\documentclass[10pt, fleqn]{article}
\usepackage[utf8]{inputenc}
\DeclareUnicodeCharacter{2212}{$-$}
\usepackage{graphicx}
\usepackage[fleqn]{amsmath}
\usepackage[version=4]{mhchem}
\usepackage{siunitx}
\usepackage{longtable,tabularx}
\usepackage{nccmath}
\usepackage{amsmath}
\usepackage[superscript]{cite}
\usepackage[margin=1in]{geometry}
\usepackage[toc,page]{appendix}
\usepackage{setspace}
\usepackage{amssymb}
\usepackage{tabu}
\usepackage{listings}
\usepackage{afterpage}
\usepackage{titlesec}
\usepackage[labelfont=bf]{caption}
\usepackage{fancyhdr}
\usepackage{pgf}
\usepackage{float}
\usepackage{subcaption}
\usepackage{hyperref}
\usepackage{cleveref}
\usepackage{textcomp}
\usepackage{pict2e}
\usepackage{wasysym}
\usepackage{multicol}
\usepackage{orcidlink}
\usepackage{lipsum}
\usepackage{url}
\usepackage{placeins}

\usepackage[parfill]{parskip}
\usepackage{enumitem}
\newenvironment{Figure}
  {\par\medskip\noindent\minipage{\linewidth}}
  {\endminipage\par\medskip}

\newcommand{\orcid}[1]{\href{https://orcid.org/#1}{\textcolor{black}{\aiOrcid}}}

\titleformat{\section}
{\normalfont\normalsize\bfseries}{\thesection}{0.5em}{}
\titleformat{\subsection}
{\normalfont\normalsize\bfseries\em}{\thesubsection}{0.5em}{}
\titleformat{\subsubsection}
{\normalfont\normalsize\bfseries\em}{\thesubsubsection}{0.5em}{}
\begin{document}

\begin{flushright}
\Large % (14 point font)

% Input the paper number: 
\textbf{SSC25-RAVII-03}
\end{flushright}
\begin{centering}      
\large % (12 point font)

% Input the title: 
\textbf{PULSE-A Mission Overview:\\
Optical Communications for Undergraduate Students}\\
\vspace{0.5cm}
\normalsize % set to 10 point font

% Input the author information:

{Logan Hanssler$^{{\orcidlink{0009-0005-6866-6583}}}$}, Seth Knights$^{{\orcidlink{0009-0009-6125-4681}}}$, Graydon Schulze-Kalt$^{{\orcidlink{0009-0003-6733-9455}}}$, Juan Ignacio Prieto Asbun$^{{\orcidlink{0009-0004-5799-9975}}}$, Robert Pitu$^{{\orcidlink{0009-0004-6637-8089}}}$,\\
Lauren Ayala$^{{\orcidlink{0009-0008-8501-671X}}}$, Rohan Gupta$^{{\orcidlink{0009-0002-1141-5853}}}$, Vincent Redwine$^{{\orcidlink{0009-0003-9713-8048}}}$, Spencer Shelton$^{{\orcidlink{0009-0006-2888-1200}}}$, Catherine Todd$^{{\orcidlink{0009-0000-4628-2785}}}$,\\
Maya McDaniel$^{{\orcidlink{0009-0001-8812-3528}}}$, Sofia Mansilla$^{{\orcidlink{0009-0001-9064-9027}}}$, John Baird$^{{\orcidlink{0009-0008-5761-7753}}}$, Mason McCormack$^{{\orcidlink{0000-0002-1463-9847}}}$, Leah Vashevko$^{{\orcidlink{0009-0000-2118-9699}}}$,\\
Tian Zhong$^{{\orcidlink{0000-0003-3884-7453}}}$\\
{University of Chicago}\\
{lhanssler@uchicago.edu}\\
\vspace{0.5cm}
Michael Lembeck$^{{\orcidlink{0000-0003-1939-6757}}}$\\
{StarSense Innovations}\\
{Champaign, IL, USA}

% Input the abstract: 
\vspace{0.5cm}
\centerline{\textbf{ABSTRACT}}
\vspace{0.3cm}
\end{centering}

Recent advances in the size, weight, and power (SWaP) requirements for space-based sensing have dramatically increased the demand for high-bandwidth downlink. However, high data rate RF transceivers still pose significant SWaP and cost restrictions, especially for university-class CubeSat missions. Optical communication may provide a solution to this challenge, enabling data transmission with order-of-magnitude rate increases over RF while being both secure and SWaP-efficient. The Polarization-modUlated Laser Satellite Experiment (PULSE-A) is a University of Chicago mission to demonstrate optical downlink at a data rate of up to 10 Mbps using circular polarization shift keying (CPolSK). PULSE-A comprises a $<$1.5U Optical Transmission Terminal, 3U CubeSat Bus, Optical Ground Station (OGS) employing an amateur telescope, and RF Ground Station (RFGS), all of which are being designed and integrated by a team of over 60 undergraduate students. The mission objective is threefold: (1) to provide hands-on educational experiences for undergraduate students, (2) to make hardware for optical communication systems more accessible via open-source design, and (3) to explore the viability and potential advantages of using CPolSK for optical downlink.

PULSE-A serves an essential educational purpose by providing University of Chicago students with the opportunity to design, build, test, and fly a spacecraft. All engineering and leadership roles on the PULSE-A Team are filled by undergraduate students from the University of Chicago Space Program (UCSP), the University’s only Registered Student Organization dedicated to engaging students in aerospace engineering projects. UCSP fulfills a unique role at the University given the absence of any conventional mechanical, electrical, or aerospace engineering programs. PULSE-A is the primary opportunity for the University of Chicago’s undergraduate students to learn and apply skills in these engineering fields, as the large majority of the mission’s hardware and software is being developed in-house. With over 100 students having worked on the mission since UCSP’s inception, PULSE-A has had a profound impact on the University’s student body. In this work, we present an overview of the mission, and we describe the PULSE-A Team’s learning-oriented approach to program management and engineering. We especially emphasize the importance of student leadership in PULSE-A’s development process and the resulting benefits for the University of Chicago community. We also highlight takeaways from the experience of founding and operating an undergraduate student-led CubeSat program, particularly regarding team organization, knowledge transfer, and collaborative learning in the context of limited prior institutional knowledge on small satellites.

\newpage
\begin{multicols*}{2}

\section*{INTRODUCTION}

\subsection*{The Need for Space-Based Optical Communications}

Over the last two decades, advances in space-based sensing have dramatically increased the amount of information, both commercial and scientific, that must be transmitted from space to ground. The proliferation of CubeSat technologies has enabled novelly small, cheap, and accessible resources for space mission development than ever before. Despite the expansion of this technology and the dramatic increase in the data acquisition capabilities of such satellites, the size, weight, and power (SWaP) constraints of CubeSat missions often pose a bottleneck to downlink communication systems\cite{Kingsbury2015}. Free-space optical communications offers a clear solution to this bottleneck, enabling order-of-magnitude higher data rates due to significantly smaller beam divergence (i.e. more efficient energy usage) and technological advances in ground-based telecommunications and fiber optics technology\cite{Hemmati_Lasercom}. The modern development of optical communication-based satellite networks opens the door not only to optical downlink, but also to high-throughput inter-satellite data relays. Even considering the more adverse effects of the atmosphere on optical links compared to RF links, the vastly increased data rates often still deliver more data to the ground with substantially less contact time.

This need for greater data throughput is especially important in scientific and academic missions with smaller budgets. In such cases, the especially high cost of large throughput RF communications systems (in particular at CubeSat SWaP) may create a fundamental barrier in collecting and completing vital research. Additionally, existing commercial off-the-shelf (COTS) optical communication terminals for CubeSats have yet to be both widely available and affordable enough to be used effectively on such missions. Thus, continued technological development is necessary to improve the reach and reliability of optical communications systems for CubeSats, and to help them impact some of the most important scientific work.

\subsection*{PULSE-A Mission Goals}

The PULSE-A mission's primary technical objective is to demonstrate space-to-ground optical communication at up to 10 Mbps using circular polarization shift keying (CPolSK). Consequently, the mission aims to provide flight qualification for custom hardware and software that facilitates polarization-based optical downlink. The custom designs are made to be open source, thus promoting accessibility for future optical communication missions. Furthermore, while optical downlink using CPolSK has been previously discussed in literature\cite{CPolSK}, it has yet to be realized in an optical downlink mission. PULSE-A will serve as a probe to explore the viability and challenges of CPolSK for optical downlink.

PULSE-A also serves an essential role in education. The mission is being developed by a team of over 60 undergraduate students at the University of Chicago. Lacking the context of a traditional engineering program, the mission’s primary objective is to serve a unique educational role for the University’s student body. By working on PULSE-A, students are able to work collaboratively in a self-motivated environment to learn skills in mechanical, electrical, software, optical, and systems engineering, as well as a number of related fields. The PULSE-A Team structure and design philosophy are oriented toward learning as the mission’s top priority.

To accomplish PULSE-A’s goals, the following key mission systems are being designed:

\begin{itemize}[leftmargin=1.25em, itemsep=0em]
    \item Custom $<$1.5U Optical Payload (acting as the transmitting terminal) and custom Optical Ground Station (OGS) (acting as the receiving terminal), each consisting of COTS optical components and custom mechanical and electrical hardware.
    \item Custom 3U Bus consisting of a custom Electrical Power System (EPS), Command and Data Handling (CDH), and Thermal Subsystem as well as COTS Attitude Determination and Control System (ADCS), RF Communications Subsystem, and Structure.
    \item Custom Radio Frequency Ground Station (RFGS) consisting of COTS components, to be established permanently at the University of Chicago.
    \item Custom Flight Software (FSW) based on NASA Goddard Space Flight Center’s open-source core Flight System (cFS) framework\cite{NASA_cFS}.
\end{itemize}

\subsection*{Concept of Operations}

\begin{figure*}
    \centering
    \includegraphics[width=\textwidth]{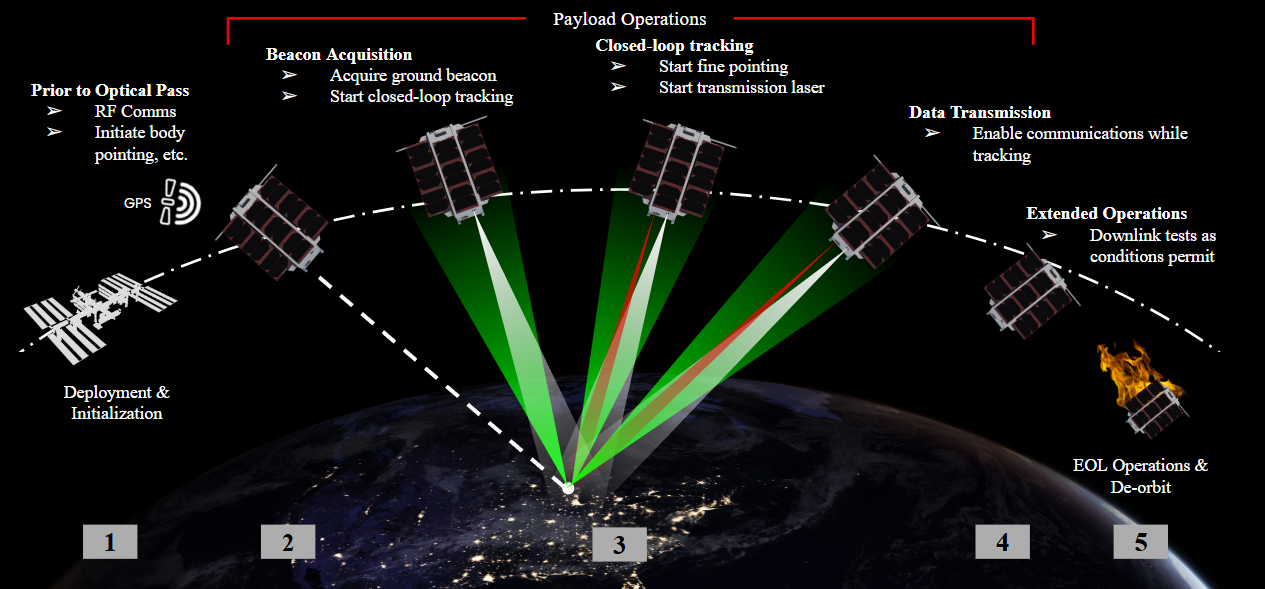}
    \captionof{figure}{Concept of Operations Visualization}
    \label{fig:conops}
\end{figure*}

The PULSE-A CubeSat will be launched through NASA’s CubeSat Launch Initiative (CSLI) no later than 2028\cite{nasa_csli} (Step 1 in Fig.~\ref{fig:conops}). Launch is currently scheduled for March 2027 and is planned for a 45º–50º inclination at 450 km–550 km.

Before any optical passes are attempted, RF communications will be established to monitor the Satellite via downlinked telemetry. PULSE-A will use its GPS module to determine its coordinates, which will then be downlinked over RF. The Ground Station will use the GPS coordinates to model the Satellite’s orbital trajectory, thus informing the Ground Station’s pointing. The Satellite will regularly downlink its GPS coordinates over RF to ensure that the Ground Station’s orbital model remains as accurate as possible. Prior to a viable optical pass, the Satellite body will point toward the OGS’s coordinates, and the OGS’s telescope will point toward the Satellite’s expected location. This expected location is calculated based on information provided via RF downlink that will be established as the Satellite passes over the horizon. (Step 2 in Fig.~\ref{fig:conops}).

If RF communication is established, and the pass is deemed to be viable, then the Satellite will attempt to point the Payload’s beacon laser toward the OGS’s telescope. Pass viability is determined by weather, the amount of time that the CubeSat will be above 30º elevation, and the availability of Ground Station operators. If the OGS successfully receives a signal from the Payload's beacon laser, then it will attempt to simultaneously point its beacon laser into the Payload’s transmission aperture. The Satellite and OGS will then enter closed-loop tracking. Using fine and coarse pointing adjustments, the OGS will attempt to center the Payload beacon on its tracking camera, which corresponds to centering the expected optical transmission beam onto the OGS’s avalanche photodiodes (APDs). The Payload will use fine pointing adjustments (combined with the Satellite body tracking the OGS) to steer the OGS beacon onto the center of its quadrant photodiode. Once the fine pointing is deemed to be sufficiently accurate, optical downlink will be attempted (Step 3 in Fig.~\ref{fig:conops}).

It is expected that mission operations will last approximately 1 year, and PULSE-A’s systems are designed such that nothing may preclude a minimum mission life of 1 year. If logistical conditions (e.g lengthened permits for Ground Station operations) and the Satellite’s expected lifetime permit, then operations may be extended to allow for more attempted optical transmissions (Step 4 in Fig.~\ref{fig:conops}). Following the end of the mission's Operations Phase, PULSE-A will enter its end of life operations and deorbit (Step 5 in Fig.~\ref{fig:conops}).

\subsection*{Major Milestones in Development}

\begin{figure*}
    \centering
    \includegraphics[width=\textwidth]{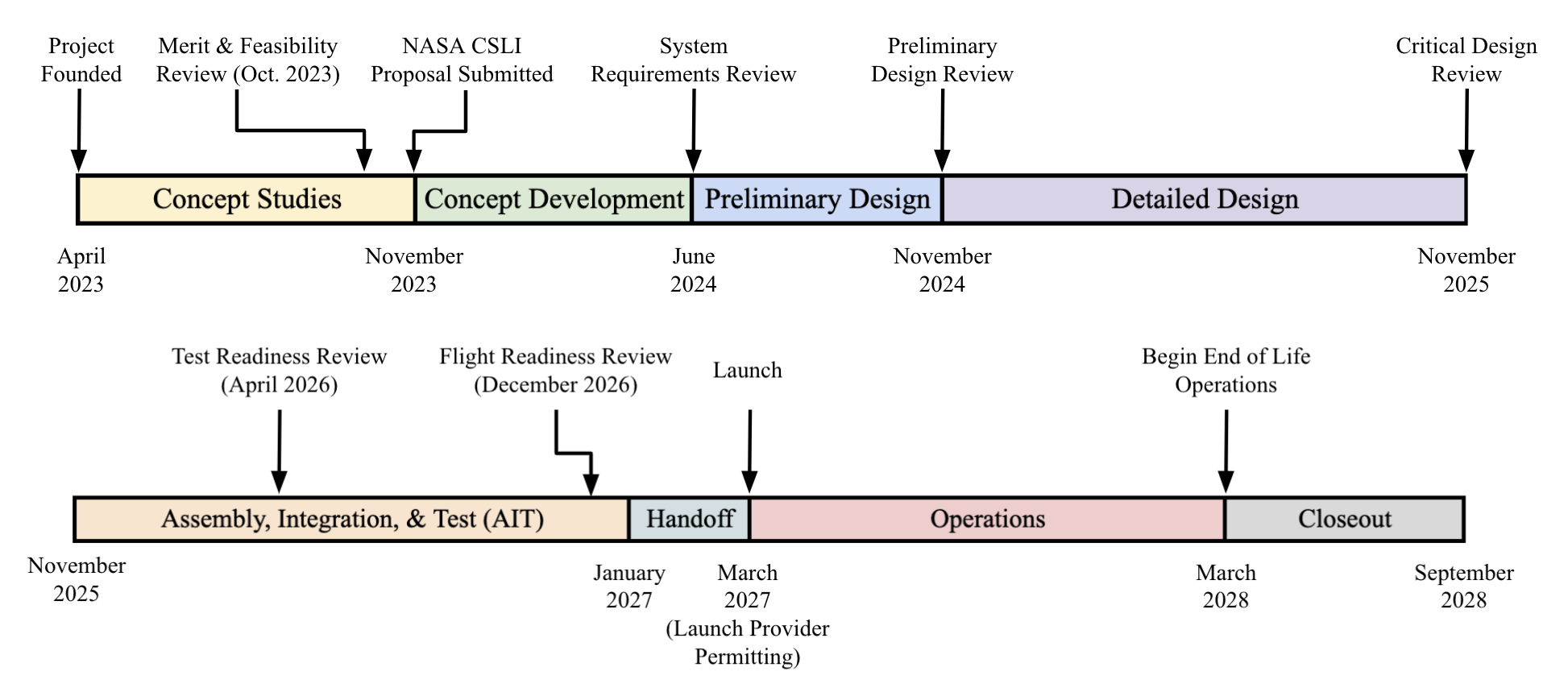}
    \captionof{figure}{Development Phase Timeline with Major Milestones Noted}
    \label{fig:timeline}
\end{figure*}

PULSE-A’s milestones and overall timeline are detailed in Fig.~\ref{fig:timeline}. Thus far, PULSE-A has successfully passed its Merit, Feasibility, System Requirements, and Preliminary Design Reviews. Additionally, the team submitted a successful proposal to NASA CSLI in November 2023, which has provided PULSE-A with a full launch sponsorship (up to \$300,000 USD). The team is currently working toward the next milestone: the Critical Design Review (CDR), after which the Assembly, Integration, and Test (AIT) Phase will begin.

The most significant milestone achieved, however, has been the team’s growth and consistent engagement. The team currently has 60 consistently contributing members, and over 100 undergraduate students have worked on PULSE-A since the mission’s inception. PULSE-A’s profound impact on the University of Chicago’s undergraduate student body is the mission’s greatest accomplishment thus far.

\section*{DESIGN OVERVIEW}

\subsection*{Design Philosophy}

PULSE-A’s design philosophy has made undergraduate education its highest priority. As a result, the vast majority of the mission is designed in-house. All of the mission's systems and the vast majority of subsystems (the Payload, OGS, RFGS, and most Bus subsystems) are fully designed, integrated, and tested by students. This has allowed PULSE-A's members to practice skills across a wide range of engineering disciplines.

A large amount of learning is required for undergraduate students to design, build, and test such custom mission systems. Therefore, the team has made an effort to prioritize documentation for all design phases. We have also provided newer members with several small targeted projects, which are not necessarily needed for the mission’s success but help them build the skills necessary to engineer PULSE-A’s systems. The mission’s timeline accounts for members’ extended learning curves, especially during transitional periods between development phases. As a result, additional time is dedicated to the Concept Studies, Detailed Design, and AIT Phases.

PULSE-A’s design emphasizes cost-efficiency to support the mission’s feasibility for an undergraduate student team. Due to several otherwise costly subsystems being developed in-house and the CubeSat prioritizing low SWaP, we have been able to develop PULSE-A at a cost of approximately \$250,000 USD.

\subsection*{Systems Engineering}

The principal mission objectives detailed in the ``PULSE-A Mission Goals" section were developed when the mission was first established. Afterward, the team created very early preliminary designs of the systems in preparation for PULSE-A’s Feasibility Review and proposal to NASA CSLI. Design work continued following the CSLI proposal’s submission until spring 2024, when system requirements were developed for the first time. The requirements were solidified in a System Requirements Document (SRD) for PULSE-A’s System Requirements Review in June 2024, and their validation methods and timeline were developed in parallel to the preliminary design until the Preliminary Design Review in November 2024. The team is currently developing the final design to satisfy the system requirements before the CDR in November 2025.

By pursuing preliminary design to such an extent before beginning to work with system requirements, PULSE-A’s members were able to quickly build familiarity with the engineering design process and the challenges associated with developing each subsystem. Given that PULSE-A is the first such student-developed mission at the University of Chicago, this learning process was necessary to gain an understanding of the engineering work to come and to obtain the level of knowledge required to implement meaningful systems engineering practices. The broad range of design and prototyping tasks undertaken from the mission’s inception also stirred excitement about PULSE-A in the student body, which proved to be crucial to the team’s rapid growth throughout 2024.

Furthermore, the early preliminary designs were not made in the absence of technical benchmarks. The team had already agreed upon a number of expected performance metrics before creating system requirements. For example, the mission’s data rate objective dictated that components chosen for the Optical Payload and OGS must enable the optical transmission to achieve an uncoded data rate of 10 Mbps. These benchmarks were not documented properly in an SRD until shortly before the System Requirements Review, largely due to the team’s insufficient systems engineering knowledge and mentorship at the time, but the design choices made prior to the SRD’s creation did conform to an implicit agreement of system requirements.

Drafting the SRD and implementing systems engineering practices around halfway through preliminary design caused relatively predictable challenges. The lack of consistent requirement documentation and systems engineering caused differences in design maturity and risk management across PULSE-A’s subsystems. Moreover, the SRD’s implementation caused the team to swiftly discover a number of conflicting design choices that had been made throughout PULSE-A’s development. Returning to the selected designs and reworking them to adhere to the mission’s requirements caused much greater consistency and design maturity across subsystems. Additionally, it allowed the team members to gain an appreciation for the iterative design process that has dominated the rest of the project’s Preliminary Design and Detailed Design Phases. This experience also allowed the team to gain an appreciation for the importance of systems engineering and documentation, which are key to ensuring adequate technical development and supporting PULSE-A’s open-source goals.

\subsection*{Optical System}

\begin{figure*}
    \centering
    \includegraphics[width=\textwidth]{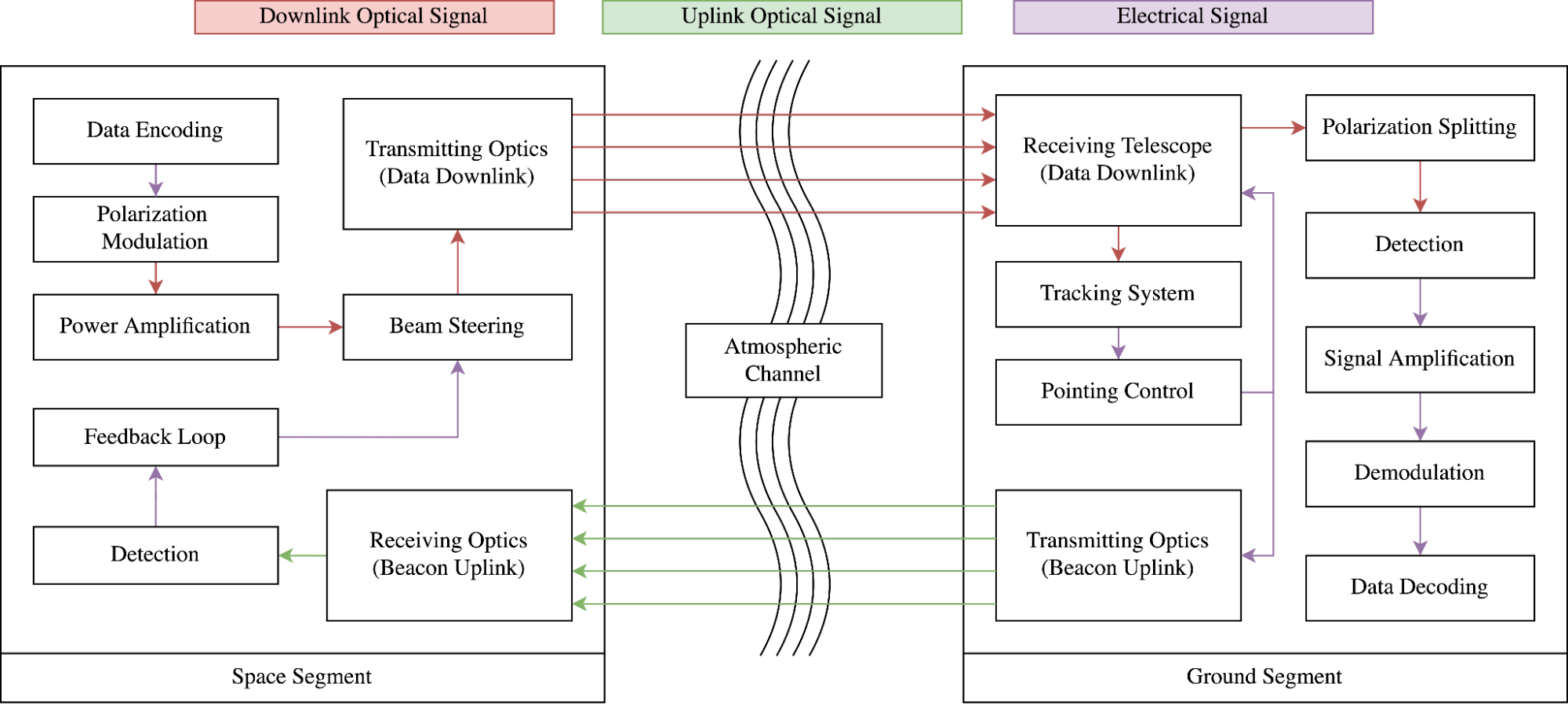}
    \captionof{figure}{Optical System Overview}
    \label{fig:optical_system}
\end{figure*}

PULSE-A’s principal technical objective of demonstrating up to 10 Mbps space-to-ground optical communications is delegated to the Satellite's Optical Payload and the OGS. The optical terminals' primary areas of development are the encoding and decoding of data into polarization-based channels and the pointing, acquisition and tracking (PAT) of highly directional optical links. To perform optical communications, the Payload transmits two laser beams: a 1550 nm transmission laser and a 638 nm beacon laser. For the Satellite to track the OGS, the OGS provides its own beacon at a wavelength of 1064 nm. This three-laser system is the core of PULSE-A’s optical design.

The Payload’s primary role is to modulate and transmit a circular polarization-keyed beam at 1550 nm, as well as to receive and track the OGS’s beacon at 1064 nm. The Payload gathers and focuses the OGS beacon light via a Keplerian beam condenser, blocks other wavelengths with a filter stack, and focuses the remaining light onto a quadrant photodiode detector. This detector supplies feedback to a fine steering mirror (FSM), which performs slight adjustments to center the beacon on the detector. The transmission path begins with two linear, orthogonally polarized seed lasers which encode data by alternately turning ON and OFF at a frequency of 1–10 MHz. This signal is then amplified to 250 mW using a random polarization erbium-doped fiber amplifier (EDFA) and passed through a quarter wave plate to convert the signal to circular polarization states. The collection and transmission paths are combined via an 1180 nm-cutoff shortpass dichroic mirror into a single optical path as they approach the FSM. This allows the transmission laser to be sent out of the Payload parallel to the OGS beacon. The Payload's beacon laser is not on the same optical path as the transmission and OGS beacon beams. Instead, the Payload's beacon laser is aligned to the Satellite's body pointing and does not have fine pointing capabilities.

The OGS must track, receive, and decode the transmitted signal from the Payload, as well as produce its own tracking beacon. The OGS utilizes a Celestron CPC1100 telescope to track, collect and condense incident light from the Payload. The Payload beacon and transmission lasers are split by the OGS. The Payload beacon is detected by a tracking camera, which uses a feedback loop to couple the transmitted CPolSK signal into the OGS’s APDs by adjusting telescope pointing and ground FSM angle. The transmission laser is split based on left- or right-handed circular polarization into two channels and converted into electrical signals by two APDs. These electrical signals are amplified, compared, and digitized to produce a string of bits. This string of bits is decoded, resulting in the completion of data transmission.

The design of a space-to-ground optical communication system provides a multitude of challenges, as developing the system requires a broad range of skills. Designing the Payload and OGS requires a large degree of optical theory, optical experimentation, optical and orbital simulation, and mechanical and electronic design. These areas and many more come into play as we tackle the challenges of designing optical communications as an undergraduate team. Substantial challenges in designing the Payload include: limits on SWaP, strong requirements to maintain polarization states, development and manufacturing of optomechanics with tight tolerances, and the development of in-house electronics and software for modulation and beacon tracking. Major challenges in developing the OGS include: developing in-house tracking algorithms and feedback loops, verification of optical link viability, development of in-house signal comparison, digitization and error correction electronics and software, and prototyping and verification of the optical system. These challenges force our team members to adapt by learning new skills which they would otherwise not encounter in their time as undergraduate students, including optical simulations, FPGA engineering, and closed-loop feedback sequences for modulation and tracking.

For an in-depth discussion of PULSE-A's Optical Payload and OGS, see Mansilla et al., 2025\cite{Mansilla2025} and Asbun et al., 2025\cite{Prieto2025}, respectively.

\subsection*{In-House Designed Bus}

\begin{figure*}
    \centering
    \includegraphics[width=0.95\textwidth]{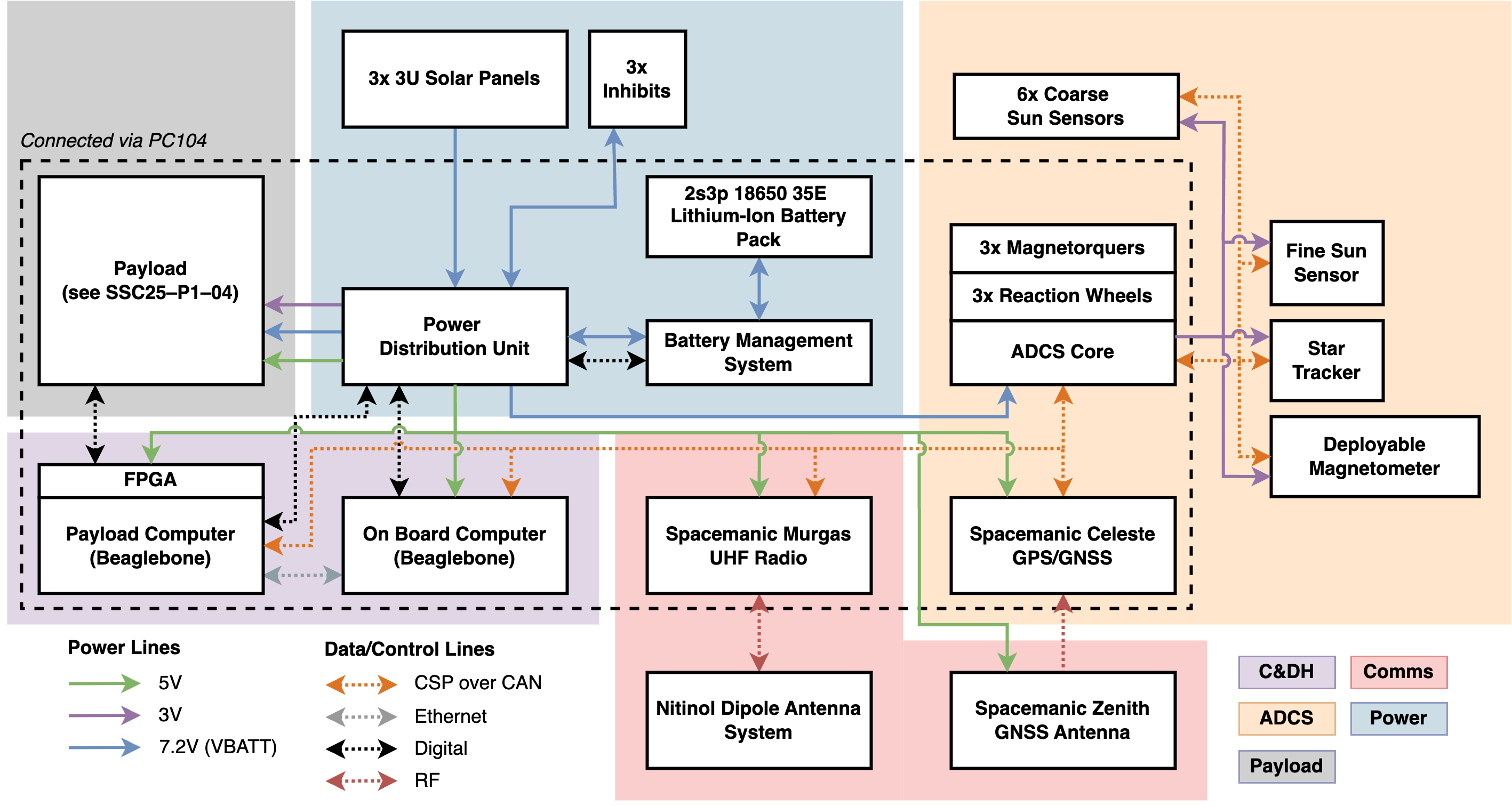}
    \captionof{figure}{Bus System Architecture}
    \label{fig:sys_arch}
\end{figure*}

PULSE-A's open-source spacecraft Bus is the backbone of the mission and has been designed in tandem with the Payload, with design driven extensively by the Payload’s pointing, power, and thermal requirements. The Bus is designed with several key principles in mind: practicality, reliability, extensibility, and affordability. The reality of university-class CubeSat development is that keeping costs low is the top priority; therefore, to practically achieve a \$250,000 USD development budget with PULSE-A’s ambitious mission, the team is developing a significant amount of hardware in-house. To feasibly achieve this while still aiming for mission success, designs are built on existing open-source packages. PULSE-A’s entire development, from proposal to launch, is slated for a less-than-three-year time frame. While cost and an emphasis on student design drive the majority of hardware being developed in-house, several components such as the ADCS, radio, and GPS are space-grade COTS to ensure reliable long-term operations. Although this increases cost, it was decided in the Concept Development Phase that in-house development of these components was infeasible considering the mission's schedule and budget.

The Bus is built on the PC/104 standard, which enables modularity and extensibility for future designs. Furthermore, the chosen software architecture makes use of a modular app-based approach. This allows for simple patching on orbit, comprehensive static and unit testing, and extensibility when reused in future missions.

The main structure of the CubeSat is a 3U aluminum frame provided by GranSystems. The frame consists of four rails compatible with PC/104 boards, four corner rails and corresponding exterior panels, and three supporting brackets for mounting. There is an additional payload box constructed in-house allowing for isolation of sensitive optical equipment. All of PULSE-A's components and mounts have been laid out 3D modeling software (with the exterior design shown in Fig.~\ref{fig:stowed_ext}-\ref{fig:deployed_ext}). 3D printing has then been used to construct a full-scale model and verify assembly procedures and fit of subsystems.

\parbox{\columnwidth}{The CDH subsystem consists of two nearly identical compute elements, dubbed the On-Board Computer (OBC) and Payload Controller, which are the Satellite’s primary control and payload control units, respectively. Although the OBC controls~all\parfillskip=0pt}

\begin{Figure}
    \centering
    \includegraphics[width = 7cm]{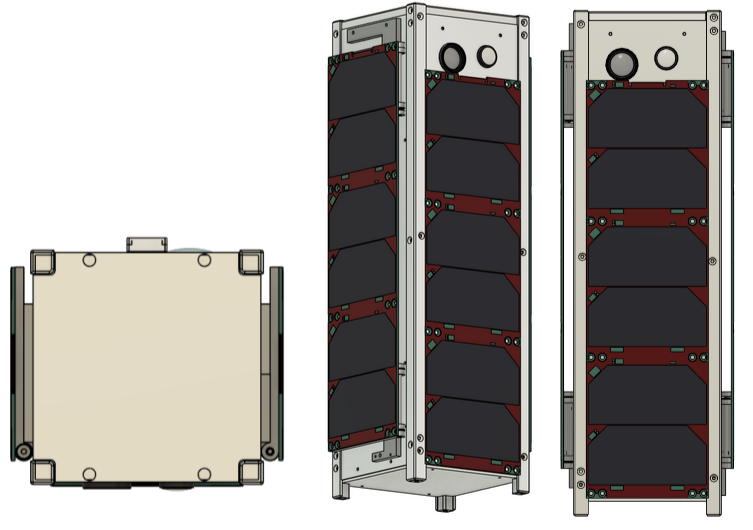}
    \captionof{figure}{Stowed CubeSat Exterior}
    \label{fig:stowed_ext}
\end{Figure}

\begin{Figure}
    \centering
    \includegraphics[width = 7cm]{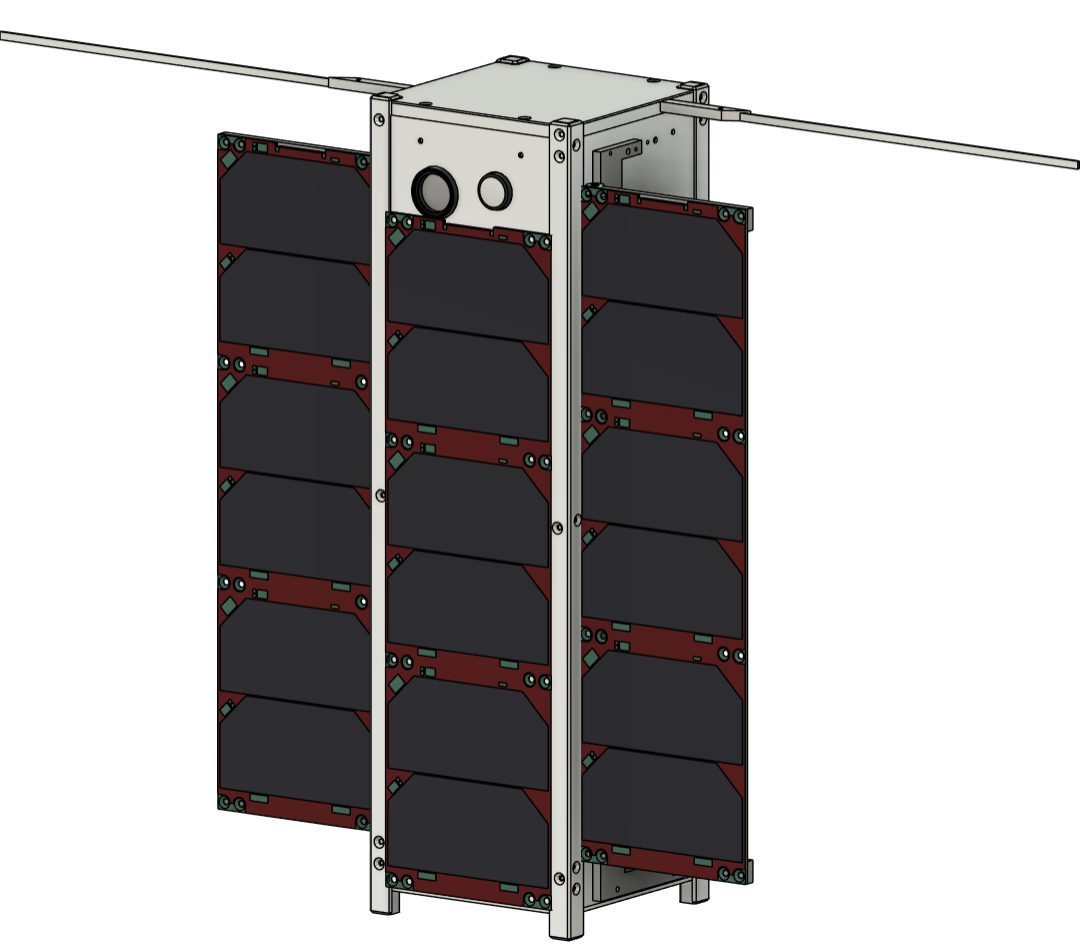}
    \captionof{figure}{Fully Deployed CubeSat Exterior}
    \label{fig:deployed_ext}
\end{Figure}

aspects of the Satellite by default, the Payload Controller serves as a redundant backup in the case of a temporary or permanent OBC failure. The BeagleBone Black, an open-source, community-driven, small form-factor embedded system with a 1 GHz ARM Cortex-A8 as its primary processing unit, serves as the main processing unit for both the OBC and Payload Controller. The Payload Controller will be heavily based on the main On-Board Computer, with slight modifications to accommodate the unique requirements of the scientific payload, including an FPGA for laser control.

The Spacemanic Murgas UHF/VHF transceiver will be used as the Satellite’s radio module. This transceiver was chosen due to its flight heritage, plug-and-play integration, low cost relative to competitors, compact size, and compatibility with the RFGS architecture. The team is currently designing a 435 MHz deployable half-wave dipole antenna fabricated from shape-memory nitinol. This antenna will be manufactured for use in the Bus.

The Power Distribution Unit (PDU) is being developed in-house by the PULSE-A Team based on extensive heritage from open-source designs from both Stanford University’s PyCubed\cite{pycubed} and the Hawai'i Space Flight Laboratory’s (HSFL) Artemis CubeSat Kit\cite{HSFL_Artemis}. In comparison to PDUs on other CubeSat missions, PULSE-A’s PDU is a “dumb” PDU—in other words, it does not contain its own compute element. Systems on board the PDU are driven by the CDH system (particularly, the OBC) via direct connections over the PC/104 stack. This design choice was made to reduce the number of independent computers on the spacecraft with different processors, necessitating different frameworks and languages, ultimately making development easier. The custom battery board is capable of fulfilling the Satellite’s high power demands while being reliable and efficient, consisting of six Samsung 35E 18650 Lithium-Ion cells in a 2s3p arrangement. The CubeSat’s power is generated by three 3U solar panels mounted in a “butterfly” or “wing” configuration, with one panel fixed to the structure and two deploying along the long axis.

The ADCS unit selected for this mission is an integrated unit provided by CubeSpace Satellite Systems with three reaction wheels, three magnetorquers, and an IMU, with an external deployable magnetometer, star tracker, coarse sun sensors, and a fine sun sensor. This system will be configured to provide better than 1° 3$\sigma$ pointing accuracy during the shaded portions of Earth orbit, where Payload operations will take place. To complement the integrated ADCS, PULSE-A is acquiring Spacemanic’s Celeste GNSS unit for the mission, allowing for highly accurate positioning of the Satellite and clock synchronization, allowing the Ground Station to improve orbital models during laser PAT regimes.

The software framework selected for the mission is NASA cFS running on a Debian-based Linux distribution with the PREEMPT\_RT patch applied. This approach affords the team a number of advantages. Debian is known to be both stable and reliable, and Linux has a history of being used as the operating system of many prior satellite missions. Applying the PREEMPT\_RT patch to the Linux kernel allows the team to take advantage of the capabilities of a real-time operating system, ensuring that critical mission events can be executed with precise timing during laser tracking while allowing for determinism when testing and debugging. cFS provides a standardized, abstracted environment to build applications within, along with a number of default apps that provide several critical services like time, tables, and file management. The standardized, abstracted nature of NASA cFS allows for all applications designed for this mission to be portable to future missions. Given that we anticipate patching software on-orbit to refine the laser tracking sequence, this modularity afforded by the application architecture allows for the team to send up individual applications to reduce the amount of transmitted data and provide a more stable way to roll out new updates. The team's FSW Department has been in close contact with the NASA cFS team at Goddard Space Flight Center to further team members’ education on FSW and NASA cFS. The collaboration between teams has allowed the PULSE-A FSW Leads to attend the first NASA cFS Symposium and the FSW Department to attend a training session hosted by Goddard.

Through leveraging commercial providers for critical and complex subsystems and in-house designs for other components, priorities in budget, flexibility in meeting mission requirements, and educational goals have been balanced throughout the Bus development process. For an in-depth discussion of PULSE-A's Bus, see Schulze-Kalt et al., 2025\cite{Schulze-Kalt2025}.

\subsection*{Radio Frequency Ground Station}

An additional major component of the PULSE-A mission is the development, construction, and commissioning of a dedicated UHF RFGS at the University of Chicago's campus. The RFGS adopts the standard Cross-Yagi architecture used by many CubeSat ground stations, but the RF chain and electronics are optimized for PULSE-A’s required high-rate GPS beaconing and command traffic. The RFGS will have the capability of full amateur-band (435-538 MHz) uplink and downlink capabilities.

Beyond serving as PULSE-A’s primary link for telemetry and command, the RFGS will be used to support future CubeSat missions developed by University of Chicago students. The RFGS will also be used as an educational tool for students, community members, and the wider Chicago amateur radio community. Planned secondary uses include SatNOGS network participation, collecting telemetry from other satellites, ham radio communications, ISS voice and packet operation, meteor-scatter experiments, and Earth-Moon-Earth (EME) communication.

\section*{TEAM OVERVIEW}

\subsection*{Establishing the Team}

The PULSE-A Team was established in April 2023 as a part of the University of Chicago Space Program (UCSP), a Registered Student Organization (RSO) at the University of Chicago committed to providing students with hands-on experience in aerospace engineering. At the time, UCSP had a small (10–20 student) membership which participated in projects such as high-power rocketry, telescope observations, and high-altitude balloon launches. UCSP’s CubeSat Laboratory was founded in December 2023 by undergraduate students seeking an accessible entrance to small satellite development. After considering various potential mission objectives, the laboratory’s members chose to work toward a space-to-ground quantum key distribution (QKD) demonstration, thus leveraging the University of Chicago’s existing resources and knowledge in quantum engineering. PULSE-A was selected as a risk reduction mission to provide flight qualification for systems capable of conducting space-to-ground optical communication using classical encryption and polarization modulation. The initially-conceived QKD mission has been dubbed PULSE-Q (Quantum) and is set to be PULSE-A’s follow-up mission.

Following PULSE-A’s founding, the mission team submitted a proposal to Call 15 of NASA’s CubeSat Launch Initiative (CSLI) in fall 2023. Throughout the proposal composition process, the team defined the mission concept and conducted successful Merit and Feasibility Reviews. This resulted in the mission’s first set of documentation, which was used to onboard new members following PULSE-A’s first large recruiting cycle in early 2024. The team’s leaders leveraged PULSE-A’s ambitious design goals and the potential launch sponsorship from NASA to rapidly expand the team with interested students, growing from less than 10 members to over 40.

In preparation for the team’s rapid growth at the start of 2024, the first hierarchical organizational structure was implemented. Five engineering departments were created: Optical Payload, Avionics Hardware and FSW, Structures and Manufacturing, Ground Station, and Systems Engineering and Integration; and an additional group was organized for PULSE-A’s Funding and Outreach. Since the team’s establishment, differing needs in successive development phases have necessitated amending the organizational structure, though the overall structure has remained relatively consistent.

\begin{figure*}
    \centering
    \includegraphics[width=\textwidth]{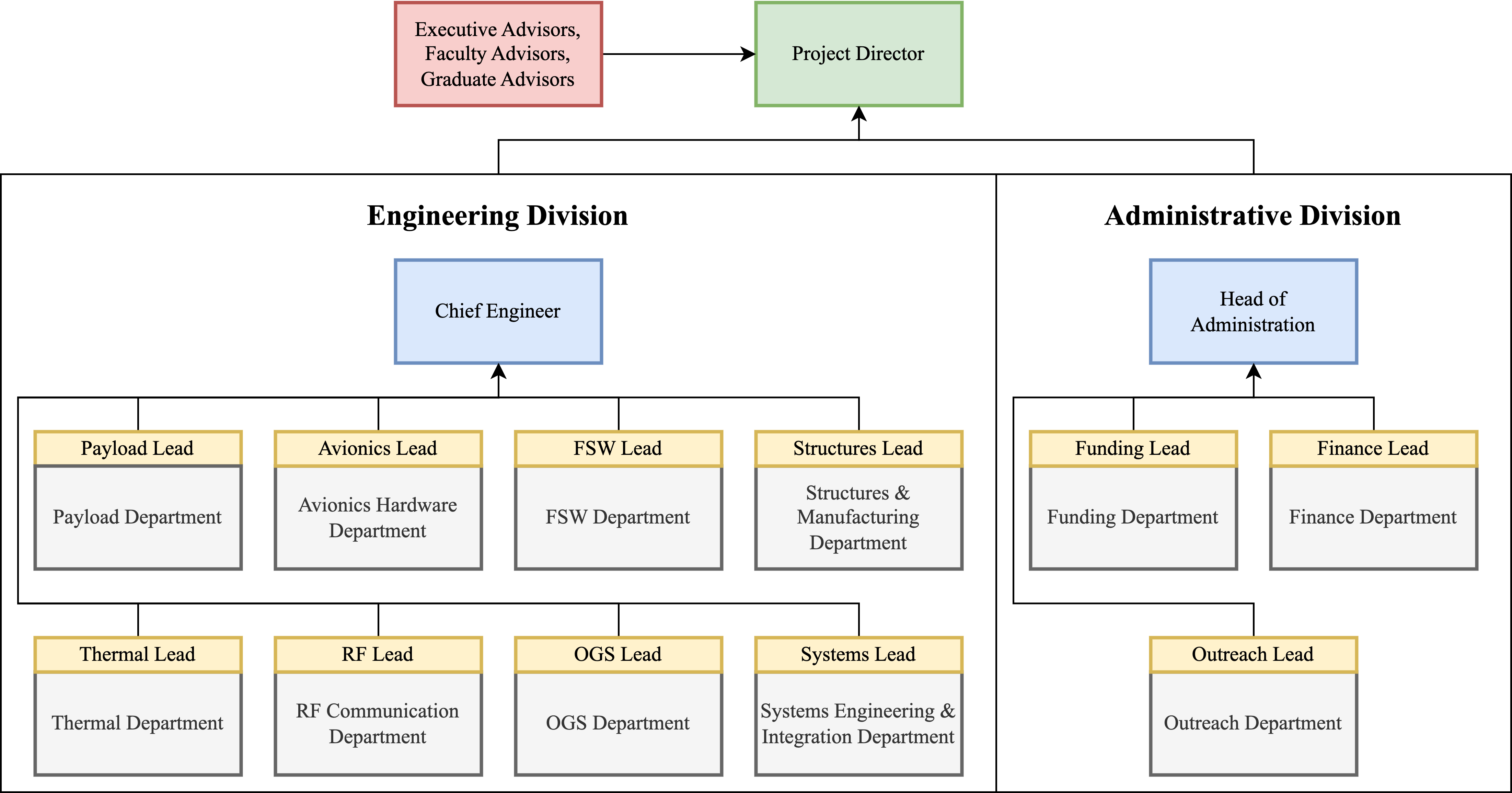}
    \captionof{figure}{Current PULSE-A Team Structure}
    \label{fig:org_chart}
\end{figure*}

\subsection*{Current Organizational Structure}

The current organizational structure (shown in Fig.~\ref{fig:org_chart}) breaks down the engineering design responsibilities very nearly by subsystem. The exceptions are the RF Communication Department, which designs the RFGS and the Satellite's RF Communication subsystem, as well as the Avionics Hardware Department, which is responsible for all other electrical Bus hardware. The engineering departments make up the team’s Engineering Division, which is paralleled by the Administrative Division. The Administrative Division consists of the Funding Department, which is responsible for grant writing and funding procurement on behalf of the project; the Finance Department, which maintains the team’s accounting and handles procurement; and the Outreach Department, which handles PULSE-A’s community outreach and guest speaker program. Each department comprises 3–10 team members led by their Department Lead. The engineering leads report to the Chief Engineer, and the administrative leads report to the Head of Administration. The Chief Engineer and Head of Administration report to the Project Director, who acts as the team lead and program manager. The Project Director, Chief Engineer, Head of Administration, Department Leads, and team members are all undergraduate students. A handful of students who previously worked in the most senior positions on the team serve as Executive Advisors to the team’s leadership. There are a small number of graduate students and faculty who serve as advisors as well.

\subsection*{Educating New Team Members}

Given the lack of traditional engineering resources at the University of Chicago, the team has almost entirely relied on independent, self-disciplined education. The need for this sort of self-study is further exacerbated by the unique knowledge required for working on laser communication systems, for which there is significantly less educational content compared to more general satellite hardware and software development. Nevertheless, the team has been successful in developing the key skills necessary to carry out such a project.  Much of this success is due to a few key aspects of the team’s recruiting and onboarding process:

\begin{itemize}[leftmargin=1.25em, itemsep=0.5em]
    \item The team encourages aspiring members to participate in other UCSP projects to gain some familiarity with the skills required of team members before applying to join. Applicants are further encouraged to take interest in a particular subsystem of the mission and self-study the requisite skills to join the team with a focus on said subsystem.
    \item The team maintains a strong body of documentation regarding subsystem design status, systems engineering, trade studies, selected components, and useful books and research papers. New members participate in a targeted program where they read and discuss these materials critically, allowing them to learn the engineering mindset and quickly catch up to speed on the mission design.
    \item After being acquainted with the mission’s documentation, new members are guided through small projects that do not usually directly contribute to the mission but allow them to practice skills they will use on the team. These projects are designed to be relatively straightforward and low-cost, thus allowing participants to focus on the key skills they are developing rather than being bogged down by logistical hurdles. Students take on these small projects in groups of 2–3. This group size allows for collaboration while being small enough that each person is pushed to take accountability for their work.
    \item While a few members are brought onto the PULSE-A Team at varying times throughout each academic year, the majority of the team’s new members join in one large recruiting cycle held at the start of the fall term. This allows the team’s new members to form a close community, sharing their educational journey through the team’s resources and designs. After completing their small projects and fully integrating with the team, each new member is also paired with a more experienced peer who shares their tasks, thus allowing new recruits to feel more comfortable in the team early on.
\end{itemize}

\subsection*{Open-Source Development}

A key aspect of the team’s founding mission was to open-source our development. During our Concept Studies and Concept Development Phases, we relied heavily on existing open-source materials to learn what the design process and teamwork should look like. PULSE-A’s development continues to rely heavily on previous open-source designs to this day. Portland State Aerospace Society’s (PSAS) open-source design documentation\cite{oresat_github}, HSFL’s Artemis CubeSat Kit\cite{HSFL_Artemis}, and Arizona State University’s Phoenix CubeSat resources\cite{asu_phoenix} were all instrumental to the PULSE-A Team finding direction shortly after its founding. As a result, one of the team’s largest goals has been to share our work as a reference for other educational CubeSat teams.

Since starting the project, we have maintained a GitHub organization\cite{github} containing copies of our proposal, all major design and requirements reviews, and major revisions of our student-designed hardware and software. While simply providing such resources can be useful to a degree, our eventual goal would be to offer supporting documentation (such as PSAS and HSFL often provide) that lowers the barrier to understanding and learning from the material. Until then, we hope that such material, alongside an openness to answering questions and collaborating with teams referencing our work, can be a resource of use for future teams like ours.

\subsection*{Student Fundraising}

The PULSE-A mission has a total budget of \$550,000 USD including launch costs, which is quite large for a student-developed mission. As a result, the team’s Funding Department has had a unique opportunity to practice fundraising in large quantities through grant applications and outreach to donors and sponsors. NASA CSLI has been the team’s largest fundraising success, guaranteeing launch sponsorship up to \$300,000 USD. Out of the remaining \$250,000 USD cost to develop the mission, the team has successfully raised \$170,000 USD thus far. Active fundraising continues, with the team’s goal being to secure all necessary funds by the CDR in advance of large-scale procurement.

\section*{KEY EDUCATIONAL TAKEAWAYS}

\subsection*{Knowledge Management}

As a student organization, the PULSE-A Team has a steady turnover due to students graduating. This means that action must be taken to ensure knowledge loss is minimized when members leave each year. The team typically prioritizes second and third year students taking on technical leadership positions, while advisory roles are reserved for fourth year students. This provides a framework for graduating students to focus on knowledge transfer while continuing students perform technical development. Additionally, documents and technical files are shared on GitHub as well as Google Drive to provide redundancy and easy access to all students. Weekly journal documentation is also provided through Notion. The team maintains comprehensive engineering documentation on these platforms.

\subsection*{Improving the Recruiting Process}

The team’s most significant recruiting challenge is finding members who are willing to dedicate the time to learning and participating in the project at an effective level. As the team has grown, we quickly learned that recruiting for particular skill sets can be time consuming and relatively unproductive. This is because the skills used to develop PULSE-A range beyond what the vast majority of students would have been exposed to before joining the team. Students who are able to successfully engage in individual projects have the requisite experience allowing them to be capable, self-sufficient learners, often engaging the most with the team. On the other hand, we have come to understand that many students have not necessarily been immersed in an environment where such individual project work was not necessarily a priority. With this in mind, our recruiting focuses mainly on interest and prior experience in other projects or research work, whether technically relevant or not. Moreover, we strive to onboard students from a diverse range of technical backgrounds (with the team representing all of the University’s STEM programs and many non-STEM programs), so that members can continue to learn from one another and further the team’s culture of collaborative and self-driven learning.

\subsection*{Student Outcomes}

PULSE-A facilitates a wide range of student outcomes both within and beyond the satellite and aerospace sectors. Graduating team members have had a variety of successful outcomes, from pursuing prestigious graduate programs to serving in the United States military. Continuing students have also earned internships at research institutions (such as NASA’s Jet Propulsion Laboratory or Argonne National Laboratory) as well as prominent engineering companies. Students have been able to utilize the novel research and design skills they gained from working on PULSE-A to leverage future employment. In addition to professional placements, PULSE-A actively encourages its members to be involved in faculty-led research. Several students have continued academic work in fields such as laser communication, optical physics, and quantum technology, contributing to both on and off-campus initiatives.

The PULSE-A Team’s Administrative Division includes students who work in capacities beyond engineering. Members of the Administrative Division have used their experience in PULSE-A’s funding, finance, and outreach efforts to find jobs and internships in finance, project management, science communication, and education.

\section*{UPCOMING DEVELOPMENTS}

\subsection*{Finalizing Design}

The team's current work is almost entirely dedicated to preparing for our upcoming CDR. This is a major turning point that takes the team from PULSE-A's design phases into the AIT Phase, where we must be prepared to rapidly produce and qualify our design for both the launch and space environments. As part of this preparation, the team will work to implement a significant portion of our hardware as a working prototype. While not explicitly necessary for a CDR, the team believes that focusing on proof-of-concept prototyping and testing is the best way to advance and finalize our designs. In addition, this will put us on track to be more competent and successful engineers during the most crucial parts of the project.

\subsection*{Further Educational Efforts}

The Fall 2025 recruiting cycle is the PULSE-A Team’s most significant upcoming opportunity to promote engineering education. It is projected that the team will onboard approximately 25 new members, with the majority being first and second year students. Due to the University of Chicago’s late start to the academic year (September 29th for the 2025–2026 year), the recruiting cycle will end in mid-October. At that point, final preparations for the CDR will be taking place. Rather than forcing the team’s new recruits to quickly become familiar with PULSE-A’s designs and haphazardly contribute to the late stages of the CDR, we plan to train new members in a targeted onboarding track. This will allow us to enact the educational plans described in the ``Educating New Team Members" section.

Furthermore, the team plans to expand the Outreach Department's educational community outreach initiatives in the 2025–2026 academic year. This will include tutorials for engineering software and design posted on social media, increased access to guest speakers from the local engineering community, and expanded partnerships with local K–12 schools. These efforts will contribute to the learning of both University of Chicago students and students in the local Chicago community.

\subsection*{Planned Follow-Up Mission}

As mentioned in the "Establishing the Team" section, PULSE-A serves as the predecessor for the PULSE-Q mission. PULSE-Q aims to be the first U.S.-based demonstration of space-to-ground QKD, marking a significant milestone in provably secure quantum communications. PULSE-Q will contribute to a rapidly evolving global landscape of satellite-based QKD efforts, which was initiated with China’s Micius satellite\cite{Micius}, launched in 2016. Although space-to-ground QKD demonstrations are currently being developed by the European Space Agency\cite{esa_qkdsat}, Singapore’s SpeQtral\cite{speqtral}, and other organizations, PULSE-Q aims to accomplish its demonstration using a smaller form factor than any other planned QKD mission by implementing a design minimally changed from that of PULSE-A. PULSE-Q would consequently serve as a technological benchmark in advancing free-space QKD systems, preparing for scalable quantum networks, and establishing the foundation for future inter-satellite QKD networks.

PULSE-Q’s Optical Payload will function as a transmission terminal, optically downlinking encryption keys from LEO to an OGS receiving terminal. We intend for PULSE-Q to employ a minimally modified version of PULSE-A’s Payload, Bus, OGS, and FSW, while PULSE-A’s RFGS will be directly repurposed for PULSE-Q. Designs will only be changed to accommodate the transmission and reception of transmission keys or to improve from lessons learned during PULSE-A’s operations where necessary. By testing early iterations of what will become PULSE-Q’s designs through the PULSE-A mission, the team will gain valuable experience and establish flight heritage. PULSE-A will also provide insight into polarization state preservation over optical downlink and PAT which will be key to ensuring PULSE-Q’s success.

\section*{CONCLUSION}

The PULSE-A Team has proven that undergraduate students can meaningfully contribute to the future of small satellite technology by designing a high-performance optical communication system in-house. PULSE-A acts as a working model for how student-led programs in aerospace engineering can thrive. With its open-source and custom-designed optical terminals, Bus, FSW, and RFGS, PULSE-A lays the groundwork for PULSE-Q and other potential missions led by future students at the University of Chicago. More than 100 undergraduate students have gained hands-on experience by working on PULSE-A, and many members have attained research positions, internships, and careers in aerospace, physics, and engineering because of their work on the mission. The PULSE-A Team is eager to provide unprecedented educational opportunities to students in UCSP throughout PULSE-A’s upcoming AIT and Operations Phases, and we hope that our work will continue to inspire students at the University of Chicago for years to come.

\section*{ACKNOWLEDGMENTS}

PULSE-A’s launch is supported through the CubeSat Launch Initiative as part of NASA’s Educational Launch of Nanosatellites (ELaNa) program, notice ID \#NNH23ZCF001. The project’s work is supported by the University of Chicago Pritzker School of Molecular Engineering, Department of Physics, Physical Sciences Division, Department of Astronomy and Astrophysics, as well as the Chicago Quantum Exchange. The project is further supported by the University of Chicago Women's Board and Select Equity, LLC. We also thank our numerous private donors who supported this work, as well as all of the students and faculty who have contributed to PULSE-A since the mission's inception.

\begin{Figure}
    \centering
    \includegraphics[width = 7.5cm]{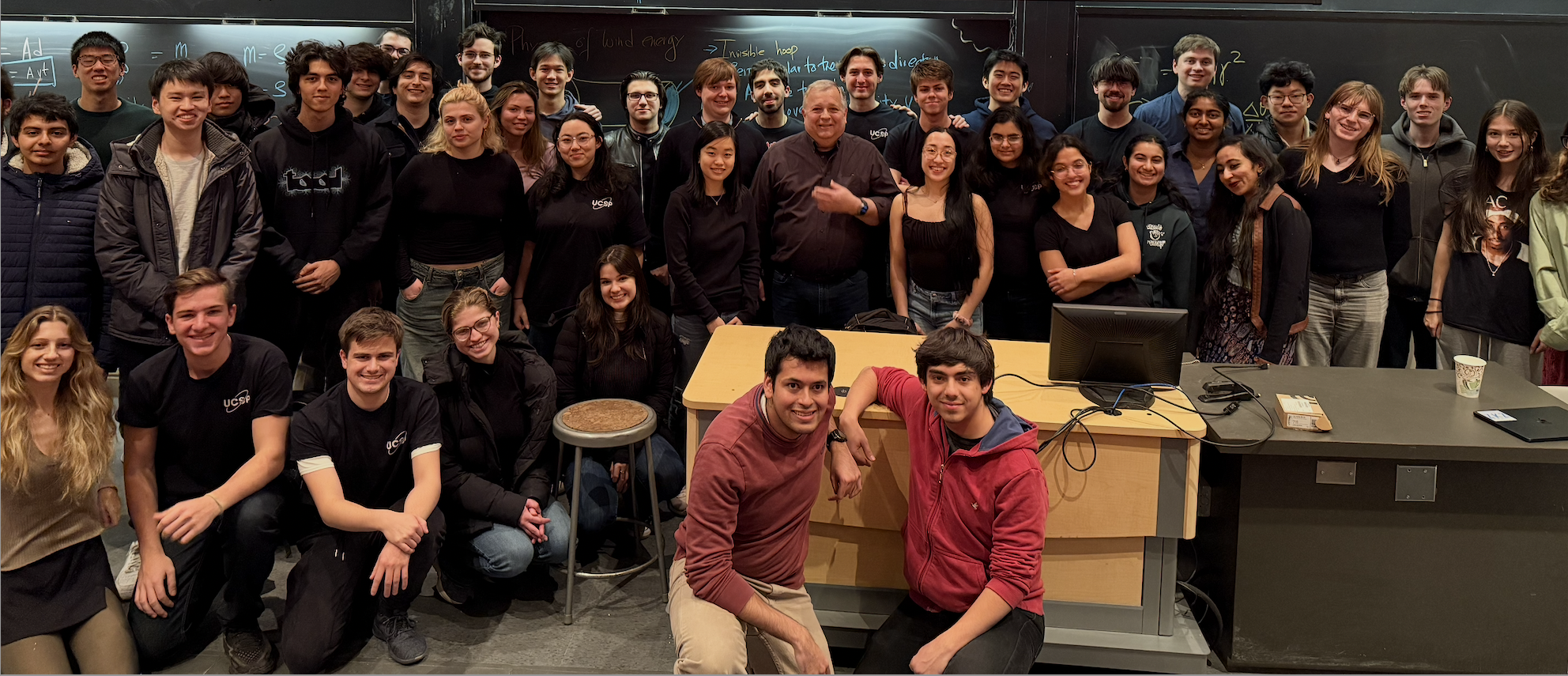}
    \captionof{figure}{PULSE-A Team in Winter 2025}
    \label{fig:pulse_team}
\end{Figure}

\section*{REFERENCES} 

\renewcommand{\refname}{}

\end{multicols*}
\end{document}